\documentstyle[twocolumn,aps,floats]{revtex}

\input psfig

\begin{document}

\twocolumn[
\hsize\textwidth\columnwidth\hsize\csname@twocolumnfalse\endcsname
\draft

\title{Scaling of the quasiparticle spectrum for d-wave
  superconductors}

\author{Steven H. Simon  and Patrick A. Lee}

\address{Department of Physics, Massachusetts
  Institute of Technology, Cambridge, MA 02139}

\date{November 18, 1996}

\maketitle
\begin{abstract}
  In finite magnetic field $H$, the excitation spectrum of the low
  energy quasiparticles in a 2 dimensional $d$-wave superconductor
  exhibits a scaling with respect to $H^{1/2}$.  This property can be
  used to calculate scaling relations for various physical quantities
  at low temperature $T$.  As examples, we make predictions for the
  scaling behavior of the finite magnetic field specific heat,
  quasiparticle magnetic susceptibility, optical conductivity tensor,
  and thermal conductivity tensor.  These predictions are compatible
  with existing experimental data.  Most notably, the thermal Hall
  coefficient $\kappa_{xy}$ measured by Krishana et al.  in YBCO is
  found to scale as $\kappa_{xy} \sim T^2 F(\alpha T/H^{1/2})$ for $T
  \lesssim 30 K$ in agreement with our predictions, where $\alpha$ is
  a constant and $F$ is a scaling function.
\end{abstract}
\pacs{PACS: 74.25.Fy, 74.72.-h, 74.25.Jb
\hspace*{115pt} {\footnotesize
    Submitted To Physical Review Letters} }

]

The scientific community has been slowly coming to a consensus that
the high $T_c$ superconductors have a $d$-wave order
parameter\cite{dwave}.  One of the major differences between these and
conventional superconductors is that $d$-wave superconductors have
gapless low energy excitations in certain directions in $k$-space,
whereas $s$-wave superconductors are gapped.  As a result, the low
temperature behavior in the $d$-wave case can be quite different from
that of conventional superconducting materials.  Thus, in order to
properly interpret experiments on these novel materials (and
eventually develop a microscopic theory), we must have a clear
understanding of the physics associated with a $d$-wave order
parameter.  In this work, we attempt to elucidate some of this physics
by deriving scaling relations obeyed by the quasiparticle energies and
eigenfunctions.  Using these relations, we are then able to deduce
scaling properties of a number of important physical quantities.

We begin our analysis with the Bogolubov\cite{general}
equations, $ {\cal H} \psi = \epsilon \psi $ where $\psi = (u,v)$ is a
Nambu 2-spinor whose components are the particle-like and hole-like
part of the quasiparticle wavefunction respectively.
Here,
\begin{equation}
  \label{eq:bdg}
  {\cal H} = \left( \begin{array}{cc} h + V - {E_{\mbox{\tiny{F}}}} & \widehat
\Delta
  \\ \widehat \Delta^* & -h^* - V  + {E_{\mbox{\tiny{F}}}}
   \end{array} \right)
\end{equation}
with ${E_{\mbox{\tiny{F}}}}$ the Fermi energy, and $h$ the kinetic part of the
effective
single particle Hamiltonian, and $V({\bf{ r}})$ is the effective disorder
potential. As a model system we will choose to work with the effective
one-particle Hamiltonian $h = \frac{({\bf{ p}} - {\bf{ A}})^2}{2 m}$ with
$m$ the electron effective mass.  Although this neglects any direct
Hartree or exchange pieces of the interaction, such pieces are thought
to be relatively unimportant except in renormalizing $m$ and $V$.  (In
this paper we have set the charge of the electron $e$, the speed of
light $c$, and Planck's constant $\hbar$ all to unity.)

In Eq. \ref{eq:bdg}, $\widehat \Delta$ is the gap operator for spin
singlet superconductivity defined as $ \widehat \Delta g({\bf{ r}}) =
\int d{\bf{ r'}} \Delta({\bf{ r}},{\bf{ r'}}) g({\bf{ r'}})$ for any
$g({\bf{ r}})$, where $\Delta({\bf{ r}},{\bf{ r'}}) = - v({\bf{ r}} -
{\bf{ r'}}) \langle \psi_{{}_{\uparrow}}({\bf{r}})
\psi_{{}_{\downarrow}}({\bf{ r'}})\rangle$ with $v$ the inter-electron
interaction.  If we rewrite $\Delta({\bf{ r}},{\bf{ r'}})$ in terms of
center of mass coordinate ${\bf{ R}} = \frac{{\bf{ r}} + {\bf{
      r'}}}{2}$ and relative coordinate ${\bf{x}} = {\bf{ r}} - {\bf{
    r'}}$, then Fourier transform with respect to ${\bf{x}}$, we can
write the gap function as $\Delta({\bf{ R}}, {\bf{ k}})$.

In this work we consider a two dimensional $d$-wave superconductor.
It is believed that this accurately represents the high $T_c$
materials. We choose to consider a gap function with pure $d_{xy}$
symmetry rather than $d_{x^2 - y^2}$ for notational simplicity. The
final results for $d_{x^2-y^2}$ are identical.  The gap function is
written as $\Delta({\bf{ R}}, {\bf{ k}}) = {\Delta_{d_{xy}}}({\bf{
    R}}) k_x k_y/({k_{\mbox{\tiny{F}}}})^2$.  Shifting back to the
coordinates ${\bf{ r}}$ and ${\bf{ r'}}$, then integrating by parts,
the gap operator can be re-expressed as
\begin{equation}
  \widehat \Delta = \frac{1}{{p_{\mbox{\tiny{F}}}}^2} \left\{p_x,
  \left\{p_y, \Delta_{d_{xy}}({\bf{ r}}) \right\} \right\}
\end{equation}
where $p_x$ and $p_y$ are the components of the momentum operator,
${p_{\mbox{\tiny{F}}}}$ is the Fermi momentum, and here the brackets represent
the
symmetrization, $\{a,b\} = \frac{1}{2} \left(a b + b a \right)$.  The
function ${\Delta_{d_{xy}}}$ is the $d$-wave order parameter used in
Ginzburg-Landau theory\cite{Ginzburg}.  We can then consider
calculating ${\Delta_{d_{xy}}}$ in an inhomogeneous system by using a
Ginzburg-Landau approach, then using ${\Delta_{d_{xy}}}$ in Eq. \ref{eq:bdg} to
find the quasiparticle spectrum.  We note that this approach is not
fully self-consistent in the sense that we will not use the derived
quasiparticle states to then recalculate the gap function.

For a homogeneous system there are gapless nodes on the Fermi surface
at the points ${\bf{ p}} = (\pm {p_{\mbox{\tiny{F}}}},0)$ and ${\bf{
    p}} = (0,\pm {p_{\mbox{\tiny{F}}}})$ where $ \widehat \Delta$
vanishes.  To study the low lying excitations near these points, we
linearize the Hamiltonian.  As an example, we consider linearizing
around the point ${\bf{ p}} = ({p_{\mbox{\tiny{F}}}},0)$.  We write $
\psi = e^{i {k_{\mbox{\tiny{F}}}} x} \tilde \psi$ such that we can
recast the Bogolubov equations as $ ( \tilde {\cal H}_0+ \tilde {\cal
  H}_1) \tilde \psi = \epsilon \tilde \psi$ where $\tilde {\cal H}_0$
is the leading linearized term
\begin{equation}
  \label{eq:newH}
  \tilde {\cal H}_0 = \left( \begin{array}{cc}
  {v_{\mbox{\tiny{F}}}}(p_x - A_x) + V &
  \frac{1}{{p_{\mbox{\tiny{F}}}}} \left\{p_y, {\Delta_{d_{xy}}}({\bf{
      r}}) \right\}
  \\  \frac{1}{{p_{\mbox{\tiny{F}}}}} \left\{p_y,
{\Delta_{d_{xy}}}^*({\bf{ r}}) \right\}
& {v_{\mbox{\tiny{F}}}}(-p_x - A_x) - V
\end{array} \right)
\end{equation}
and $\tilde {\cal H}_1$ is the remaining piece
\begin{equation}
  \tilde {\cal H}_1 = \left( \begin{array}{cc} h & \widehat \Delta
  \\ \widehat \Delta^* & -h^* \end{array} \right)_,
\end{equation}
where ${v_{\mbox{\tiny{F}}}} = {p_{\mbox{\tiny{F}}}}/m$ is the Fermi velocity.

For sufficiently small energy excitations, $\tilde {\cal H}_0$ is much
greater than $\tilde {\cal H}_1$ and it will be a reasonable
approximation to neglect $\tilde {\cal H}_1$.  To determine when this
is a good approximation, we consider the homogeneous case of
${\Delta_{d_{xy}}}$ a real constant with ${\bf{ A}} = 0$, and $V=0$.
We then find that $\tilde {\cal H}_0$ is just the Dirac Hamiltonian
for massless fermions in two dimensions, and thus has a conical linear
spectrum of quasiparticles $\epsilon_{{\bf{ p}}} = \pm {{\cal
    E}}({\bf{ p}})$ with ${{\cal E}}({\bf{ p}}) =
\sqrt{({v_{\mbox{\tiny{F}}}} p_x)^2 + ({\Delta_{d_{xy}}}
  p_y/{p_{\mbox{\tiny{F}}}})^2}$.  Note that the conical spectrum is
highly anisotropic since ${v_{\mbox{\tiny{F}}}} \gg
{\Delta_{d_{xy}}}/{p_{\mbox{\tiny{F}}}}$.  For excitations at
temperature $T$, the typical momenta are $p_x \sim
T/{v_{\mbox{\tiny{F}}}}$ and $p_y \sim T
{p_{\mbox{\tiny{F}}}}/{\Delta_{d_{xy}}}$.  The largest term in $\tilde
{\cal H}_1$ is then the term $p_y^2/(2 m)$ which would be on order
${E_{\mbox{\tiny{F}}}} (T/{\Delta_{d_{xy}}})^2$.  For YBCO,
photoemission spectroscopy\cite{photo} indicates that
${E_{\mbox{\tiny{F}}}} \approx 3,000 K$, and ${\Delta_{d_{xy}}}
\approx 450K$.  In YBCO, we also note that the Fermi surface is not
circular, but is somewhat flattened at the nodes (more square-like
with rounded corners).  This means we should really use an effective
mass $m_y$ in the $p_y^2/(2m)$ term of $\tilde {\cal H}_1$ which
lowers the energy scale of $\tilde {\cal H}_1$ by another factor of
perhaps two or three.  Thus, we estimate that $\tilde {\cal
  H}_1/\tilde {\cal H}_0 \approx T/(200K)$, so that the condition
$\tilde {\cal H}_0 \gg \tilde {\cal H}_1$ may be well satisfied at
temperatures as high as $30K$.

We note that in Eq. \ref{eq:newH} a nonzero $A_x$ acts as a scalar
potential for the quasiparticles.  If we consider the case of a
uniform superfluid velocity, we can choose a gauge where
${\Delta_{d_{xy}}}$ is a real constant and ${\bf{ A}}$ is proportional
to the velocity (i.e., London gauge).  In this case, $A_x$ in Eq.
\ref{eq:newH} acts as a scalar potential to yield a Doppler shifted
spectrum $\epsilon_{{\bf{ p}}} ={v_{\mbox{\tiny{F}}}} A_x \pm {{\cal
    E}}({\bf{ p}})$.  Of course, were we to consider quasiparticles
near the opposite $(-{p_{\mbox{\tiny{F}}}},0)$ node, we would have a
spectrum $\epsilon_{{\bf{ p}}} = - {v_{\mbox{\tiny{F}}}} A_x \pm
{{\cal E}}({\bf{ p}})$ .  As pointed out by Volovik\cite{Volovik}, in
a magnetic field, screening currents exist with a typical velocity
proportional to $H^{1/2}$ so that in a semiclassical approximation,
the density of states at zero energy is proportional to $H^{1/2}$.

Let us now apply a magnetic field $H$ perpendicular to the plane of
the sample to create a vortex lattice such that the phase of the gap
${\Delta_{d_{xy}}}$ twists a full $2 \pi$ as we go around each vortex.
(Since the screening length is very long, we can assume $H$ is
homogeneous).  The distance between the vortices is proportional to
the magnetic length $l_H \sim H^{-1/2}$.  We now claim that at low
$T$, to a very good approximation, the Hamiltonian $\tilde {\cal H}_0$
has a simple scaling form that we write (in a slight abuse of
notation) as $\tilde {\cal H}_0^H({\bf{ r}}) = [H/H_0]^{\frac{1}{2}}
\tilde {\cal H}_0^{H_0}({\bf{ r}} [H/H_0]^{\frac{1}{2}})$.  In other
words, if we can find the eigenvectors $\tilde \psi_n^{H_0}({\bf{
    r}})$ and eigenenergies $\epsilon_n^{H_0}$ of the Hamiltonian
$\tilde {\cal H}_0$ in field $H_0$, then the eigenenergies and
eigenvectors in field $H$ can be written as
\begin{eqnarray} \label{eq:sca1}
  \tilde \psi_n^H({\bf{ r}}) &=& \tilde \psi_n^{H_0}({\bf{ r}}
[H/H_0]^{\frac{1}{2}} ) \\ \epsilon_n^H &=& [H/H_0]^{\frac{1}{2}}
\epsilon_n^{H_0}. \label{eq:sca2}
\end{eqnarray}
The first of these equations is the statement that the functional form
of the eigenvector scales as the vortex lattice, whereas the second is
just a reflection of the Hamiltonian being linear in momentum.  In
order to demonstrate that these scaling properties hold, we consider
each term in $\tilde {\cal H}_0$ individually.  It easy to show that
the vector potential in a field $H$ can be written in a scaling form
${{\bf{ A}}}_H({\bf{ r}}) = [H/H_0]^{\frac{1}{2}} {\bf{
A}}_{H_0}({\bf{ r}} [H/H_0]^{\frac{1}{2}})$, and similarly, ${\bf{
p}}$ must scale as the inverse of the characteristic length $l_H$ so
that ${{\bf{ p}}}^H = [H/H_0]^{\frac{1}{2}} {{\bf{ p}}}^{H_0}$.  Thus
we need only examine $V$ and ${\Delta_{d_{xy}}}$.

We first consider the scaling of ${\Delta_{d_{xy}}}$.  In order to
have the desired scaling of $\tilde {\cal H}_0$, we must have
${\Delta_{d_{xy}}}^H({\bf{ r}}) = {\Delta_{d_{xy}}}^{H_0}({\bf{ r}}
[H/H_0]^{\frac{1}{2}})$.  This is simply the statement that, like the
wavefunction $\tilde \psi$, the functional dependence of
${\Delta_{d_{xy}}}$ on position scales with the vortex lattice.  This
assumption is clearly not correct near the vortex cores where there is
some fixed width to the core that is relatively independent of field.
However, away from the cores, for $T \ll T_c$, the magnitude of
${\Delta_{d_{xy}}}$ is fixed and only the phase varies.  For
$|{\Delta_{d_{xy}}}|$ fixed, it can be shown that the function form of
the phase that minimizes the Ginzburg-Landau free energy has the
correct scaling properties.  Thus, for $T\ll T_c$, we expect that
Hamiltonian will scale to a very good approximation.

We note that in general, currents in $d$-wave superconductors can
introduce some small amount of $s$-wave component ($\Delta_s$) to the
order parameter\cite{sdmixing}.  It can also be shown that the above
scaling laws hold to linear order in $\Delta_s/{\Delta_{d_{xy}}}$, so that for
${\Delta_{d_{xy}}} \gg \Delta_s$ the scaling laws (Eqs. \ref{eq:sca1} and
\ref{eq:sca2}) are preserved\cite{Melater}.

Finally, we turn to consider the disorder term $V$.  For Gaussian
delta-function correlated disorder such that $\langle V \rangle = 0$,
and $\langle V({\bf{ r}}) V({\bf{ r'}}) \rangle = V_0 \delta({\bf{ r}}
- {\bf{ r'}})$, the disorder does not define a length scale so that
given a realization of disorder $V({\bf{ r}})$, another configuration
$V'({\bf{ r}}) = [H/H_0]^{\frac{1}{2}} V({\bf{ r}}
[H/H_0]^{\frac{1}{2}})$ is equally likely.  In other words, the
disorder term (in an ensemble average) has the proper scaling
properties to preserve Eqns.~\ref{eq:sca1} and \ref{eq:sca2}.  For
more general disorder with a nonzero correlation length, these scaling
laws will no longer hold precisely.  We note, however, that since
there is no Anderson's theorem\cite{general} for d-wave
superconductors, the introduction of disorder will reduce the overall
value of the gap and thereby reduce the maximum temperature at which
the condition $\tilde {\cal H}_0 \gg \tilde {\cal H}_1$ is satisfied.

Neglecting disorder once again, ${\Delta_{d_{xy}}}$ and ${\bf{ A}}$
can both be considered to be periodic functions with the periodicity
of the vortex lattice\cite{Melater}.  Due to this periodicity, the
eigenstates can be divided into Brillouin zones with one band of
excitations per zone.  The first zone should have a maximum $k$-vector
of approximately $|{\bf{ k_{max}}}| \approx l_H^{-1}$, with $l_H$ the
magnetic length.  The number of different zones with momentum less
than some $k$ is roughly $(k/k_{max})^2$.  The typical energy scale of
an excitation of wavevector $k$ is $k \sqrt{{\Delta_{d_{xy}}}
  {v_{\mbox{\tiny{F}}}}/{p_{\mbox{\tiny{F}}}}}$.  Thus, the typical
energy $E_n$ of the $n^{th}$ band is given roughly by $ E_n \sim
k_{max} \sqrt{ n {\Delta_{d_{xy}}}/m} \sim \sqrt{n \omega_c
  {\Delta_{d_{xy}}}}$ where $\omega_c = B/m$ is the cyclotron
frequency.  It is then convenient to define the dimensionful constant
$\alpha = \sqrt{m/{\Delta_{d_{xy}}}}$ such that $ E_n^2 \approx n
\alpha^{-2} B$.  Finally, it will be useful to define the
dimensionless parameter $ x = \alpha T/H^{1/2} $ which is roughly the
number squared of bands that are considerably occupied at temperature
$T$.  For YBCO, $\alpha \approx .05 \mbox{Tesla}^{1/2}/K$.

Using the above described scaling laws we can extract a number of
important statements about physical quantities.  As a first example we
examine the specific heat.  We write the energy as
\begin{eqnarray}
  U &=& \sum_n \epsilon^H_n f(\epsilon^H_n/T) \nonumber \\ &=& [H/H_0]^{1/2}
  \sum_n \epsilon^{H_0}_n f(\epsilon^{H_0}_n[H/H_0]^{\frac{1}{2}} /T)
  \label{eq:sum1}
\end{eqnarray}
where $f$ is the Fermi function.  The volume $\nu$ of the system here
scales as $l_H^2 \sim 1/H$ so that $\nu = \nu_0 [H_0/H]$.  Thus, the
energy density can be written as $U/\nu = H^{3/2} F_U(\alpha
T/H^{1/2})$ where $F_U$ is some scaling function that we can write
down in terms of eigenenergies, but can not completely evaluate
without fully diagonalizing $\tilde {\cal H}_0$.  Here we have used
the fact that the sum in Eq.  \ref{eq:sum1} is only a function of the
dimensionless quantity $x = \alpha T/H^{1/2}$.  Note that here and
elsewhere in this paper, we will assume that we are sufficiently far
below $T_c$ that the magnitude of the gap does not change much with
$T$.  Differentiating to obtain the electronic specific heat per unit
volume we find that
\begin{equation}
  \label{eq:cv}
  C_v = T H^{\frac{1}{2}} F_C(\alpha T/H^{\frac{1}{2}})
\end{equation}
where $F_C$ is again some unknown scaling function.  We note that this
scaling form does not include contributions to the specific heat from
normal state electrons in the vortex cores.  These contributions,
however, are thought to be small\cite{Volovik}.  As discussed above,
in a semiclassical approximation, the density of states at zero energy
is proportional to $H^{1/2}$ so that the specific heat is proportional
to $T H^{1/2}$ at low temperatures as predicted by
Volovik\cite{Volovik}.  Thus the function $F_C$ should be a constant
at small argument.  Deviations from this constant should first occur
when we start to fill more than one band.  This happens at $x \approx
1$, or at $(T/H^{1/2}) \approx 20 \,\,\, K/\mbox{Tesla}^{1/2}$.

Similarly, starting with the free energy $F = U - TS$, where $S =
\sum_n \left(f_n \ln f_n + (1-f_n) \ln (1-f_n)\right)$ and $f_n =
f(\epsilon_n/T)$ we can write the free energy per unit volume in the
scaling form $F/\nu = H^{3/2} F_F(T/H^{1/2})$.  We then conclude that
the quasiparticle magnetic susceptibility per unit volume scales as $
\chi = d^2 (F/\nu)/dM^2 = \frac{1}{T} F_\chi(\alpha T/H^{\frac{1}{2}})
$ with $F_\chi$ an unknown scaling function.  Since there is no
crossing of states through the Fermi level as we change magnetic
field, we do not predict any de Haas-van Alphen oscillations.
However, there may be oscillatory contributions to the susceptibility
from the condensed fraction and the normal vortex cores that we do not
consider here\cite{Norman}.

We now turn to consider electrical and thermal transport properties.
We first define\cite{Mahan} the charge velocity operator ${\bf{
    v}}^{(1)} \equiv i [{\cal H},{\bf{ r}} \sigma_z]$ and the thermal
velocity operator ${\bf{{v}}}^{(2)} \equiv i [ {\cal H}, \{ {\cal H},
{\bf{ r}} \} ]$.  Operating on a state $\psi = e^{i
  {k_{\mbox{\tiny{F}}}} x} \tilde \psi$ near the node at
$({p_{\mbox{\tiny{F}}}},0)$, we find
\begin{eqnarray}
  {\bf{ v}}^{(1)} e^{i {k_{\mbox{\tiny{F}}}} x} \tilde \psi &=& e^{i
    {k_{\mbox{\tiny{F}}}} x} [\tilde {\cal H}_0, \{\sigma_z , {\bf{
      r}} \}] \tilde \psi + \mbox{smaller terms}
  \\ {\bf{ v}}^{(2)} e^{i {k_{\mbox{\tiny{F}}}} x} \tilde \psi &=& e^{i
{k_{\mbox{\tiny{F}}}} x} [\tilde
  {\cal H}_0, \{\tilde {\cal H}_0, {\bf{ r}} \}] \tilde \psi +
  \mbox{smaller terms}
 \label{eq:ve}
\end{eqnarray}
where the smaller terms are typically smaller by order $\tilde {\cal
  H}_1 / \tilde {\cal H}_0$.  It is then easy to see from this form
that the operator ${\bf{ v}}^{(2)}$ scales as $H^{1/2}$ whereas the
operator ${\bf{ v}}^{(1)}$ scales as $H^0$.

We now use the Kubo formula to write the generalized response function
at frequency $\omega$ by\cite{Mahan}
$$ L_{ij}^{ab} = \frac{T}{\nu} \sum_{nm} \frac{\langle n | {\bf{
      v}}^{(a)}_i | m \rangle \langle m | {\bf{ v}}^{(b)}_j | n
  \rangle f(\epsilon_n/T, \epsilon_m/T)}{(\epsilon_n - \epsilon_m -
  \omega - i 0^+)(\epsilon_n - \epsilon_m + i 0^+)}
$$
where $f$ is the thermal occupation factor, $\nu$ is the volume of the
system, the indices $i$ and $j$ take the values
${\bf{\hat{x}}}$ and ${\bf{\hat{y}}}$,
and the indices $a$ and $b$ take the values $1$ and $2$ (for
charge and heat transport respectively).  Noting that the volume of
the system scales as $H^{-1}$ and the energies all scale as $H^{1/2}$
we immediately obtain the two parameter scaling law $ L_{ij}^{ab} \sim
T^{a+b-1} F_{ij}^{ab}(\alpha T/H^{1/2}, \alpha \omega/H^{1/2})$, where
$F_{ij}^{ab}$ is again some scaling function that we will not
evaluate.  The real part of the optical conductivity tensor is defined
as $ \mbox{Re}[\sigma_{ij}] = \frac{1}{T} \mbox{Re}[L^{11}_{ij}]$
which immediately yields a two parameter scaling law for the optical
conductivity.
\begin{equation}
 \mbox{Re} [\sigma_{ij}] \sim
F_{ij}^{11}(\alpha T/H^{1/2}, \alpha \omega/H^{1/2}).
\end{equation}
It should be noted that in this Kubo formula calculation the response
of the superfluid fraction has been neglected.  This then does not
include, for example, the response of the system due to the motion of
vortices\cite{OpticalHall}.

\begin{figure}
  \centerline{\psfig{file=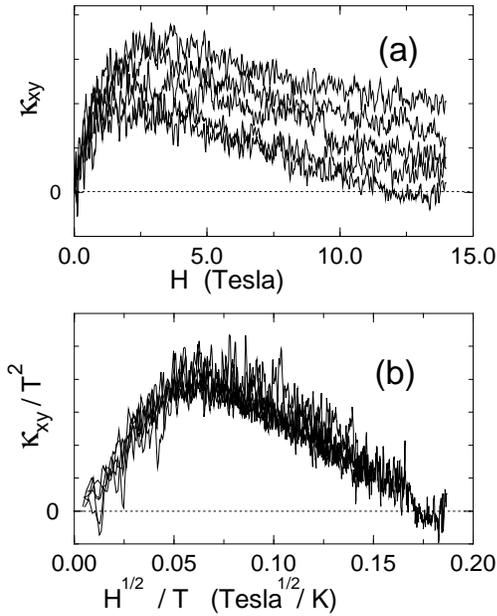,width=.8\linewidth}}
  \caption{Thermal Hall data from reference \protect \onlinecite{Krish2}.
(a) Thermal Hall transport coefficient $\kappa_{xy}$ plotted against
external magnetic field $H$ at temperatures (from bottom to top) $20K,
22.5K, 25K, 27.5K$, and $30K$.  For technical reasons data has not yet
been taken below 20K.  (b) Collapse of these five curves according to
the scaling law shown in Eq.  \protect \ref{eq:scale2}.  Note that the
characteristic scale of $H^{1/2}/T$ is approximately $.05 \,\,\,
\mbox{Tesla}^{1/2}/K$.}
\label{fig:krish}
\end{figure}

We now turn our attention to the DC ($\omega = 0$) thermal
conductivity tensor $\kappa$, defined as the matrix that relates the
heat current ${\bf{ j}}_q$ to the temperature gradient via ${\bf{
    j}}_q = \kappa \nabla T$.  We note that experimentally, a large
part of the diagonal components of this tensor are due to phonon
transport of heat.  However, the Hall (off diagonal) component of this
tensor should be completely electronic in origin\cite{Krishana}.
Note, that when calculating $\kappa$, one must usually take into
account the effect of the thermoelectric coefficient $L^{12}$.
However, here we can neglect that contribution, since there is never
any voltage in the superconducting state.  Thus, we have $\kappa_{ij}
= \frac{1}{T^2} L_{ij}^{22}$, and we obtain the naive scaling law
$\kappa_{ij} \sim T F^{22}_{ij}(\alpha T/ H^{1/2})$.  Although this is
indeed the correct scaling form for the (electronic part of the)
diagonal component of the tensor, it is not correct for the Hall
component.  It can in fact be shown that the scaling function
$F_{xy}^{22}$ here is precisely zero due to the particle-hole symmetry
inherent in the linearized Hamiltonian $\tilde {\cal H}_0$.  This
result is very easy to understand.  Imposing a heat source on one side
of the system excites many particles and holes.  Both particles and
holes diffuse in the direction of the heat sink.  In a magnetic field,
the particles curve one way and the holes curve the other way.  Thus,
when there is particle-hole symmetry, there is no net Hall transport
of heat.

The proof that the linearized Hamiltonian yields $\kappa_{xy}=0$ is a
little bit involved.  The particle-hole symmetry is expressed
mathematically by saying that given an eigenpair $\epsilon_n, \tilde
\psi_n$ satisfying $\tilde {\cal H}_0 \tilde \psi_n = \epsilon_n
\tilde \psi_n$, it can be shown that there exists another eigenvector
$\tilde \psi'_n = \sigma_y \tilde \psi_n^*$ with the same eigenvalue
also satisfying $\tilde {\cal H}_0 \tilde \psi'_n = \epsilon_n \tilde
\psi'_n$ where $\sigma_y$ is the usual Pauli spin matrix.  Using this
symmetry and adding up the contributions to $\kappa_{xy}$ from all
four nodes on the Fermi surface, it can be shown that $\kappa_{xy}$
vanishes.  It should be noted that $\kappa_{xy}$ remains zero even
when we include the smaller terms in from Eq.  \ref{eq:ve}.  We also
note that this mechanism does not force $\sigma_{xy} = 0$.  Further
details of this calculation will be published elsewhere\cite{Melater}.

In order to find a nonzero $\kappa_{xy}$, we must break the
particle-hole symmetry by including the contributions from $\tilde
{\cal H}_1$.  As mentioned above, at low enough $T$, we have $\tilde
{\cal H}_1 \ll \tilde {\cal H}_0$ so that this term can be treated
perturbatively.  Inclusion of this term then shifts the eigenenergies
via $\epsilon_n \rightarrow \epsilon_n + \delta \epsilon_n$ and the
eigenstates $|n\rangle \rightarrow |n\rangle + \delta |n \rangle$.  To
lowest order these shifts are given by the usual expressions from
perturbation theory which then obtain the following scaling forms
\begin{eqnarray}
  \delta \epsilon^H_n &=& \langle n^H | \tilde {\cal H}_1 | n^H
  \rangle \nonumber  = [H/H_0]^{1/2} \delta \epsilon^{H_0}_n
\\ \delta |n^H \rangle & = & \sum_m \frac{|
  m^H \rangle \langle m^H | \tilde {\cal H}_1 | n^H
  \rangle}{\epsilon_n^H - \epsilon_m^H} \nonumber = [H/H_0]^{1/2}
\delta |n^{H_0} \rangle.
\end{eqnarray}
Note that both of these corrections scale as $H^{1/2}$.  Including
these first order corrections into the Kubo formula and expanding we
find these correction terms give a leading contribution to the thermal
Hall conductivity that scales as
\begin{equation}
  \label{eq:scale2}
  \kappa_{xy} \sim T^2 F_{\kappa_{xy}}(\alpha T/H^{1/2})
\end{equation}
with $F_{\kappa_{xy}}$ again some scaling function.  As shown in Fig.
1, experimental results of reference \onlinecite{Krish2} do indeed
show this scaling form at temperatures below $30 K$ (For technical
reasons data has not yet been taken at temperatures below $20 K$).  As
discussed above, the characteristic scale for features in the function
$F_{\kappa_{xy}}$ (i.e., where the curve becomes nonlinear) should be
seen at $x \approx 1$ or $H^{1/2}/T \approx .05 \,\,\,
\mbox{Tesla}^{1/2}/K$, which is in good agreement with experiment.

In conclusion, we have found that the scaling properties of the
quasiparticle spectrum in $d$-wave superconductors provides a very
general and powerful tool for analyzing various physical quantities.

It is a pleasure to acknowledge very helpful conversations with K.
Matveev, D. K. K. Lee, and particularly L. S. Levitov.  The authors
are greatly indebted to N. P. Ong and K.  Krishana for sending us
their experimental data prior to publication.  This work was supported
by NSF Grant Number DMR-9523361.

\end{document}